# Band Renormalization in a Metal–Organic Framework/Au(111) Triple-Lattice Heterostructure

Xiaoqing Yuan[1,2,3,†], Shaoze Wang[2,3,4,†,*], Xiaoyue He[5], Zhecheng Sun[3,4], Zihao Wang[6], Fei Song[7], Lei Sun[2,3,4,*]

[1]Department of Physics, Fudan University, Shanghai 200433, China

[2]Department of Physics, School of Science and Research Center for Industries of the Future, Westlake University, Hangzhou 310030, Zhejiang Province, China

[3]Institute of Natural Sciences, Westlake Institute for Advanced Study, Hangzhou 310024, Zhejiang Province, China

[4]Department of Chemistry, School of Science and Research Center for Industries of the Future, Westlake University, Hangzhou 310030, Zhejiang Province, China

[5]Songshan Lake Materials Laboratory, Dongguan 523808, Guangdong Province, China

[6]School of Physics, Hunan Key Laboratory for Super-Microstructure and Ultrafast Process, Central South University, Changsha 410083, Hunan Province, China

[7]Shanghai Synchrotron Radiation Facility, Shanghai Advanced Research Institute, Chinese Academy of Sciences, 20100, Shanghai, China

[†]These authors contributed equally to this work.

*Email: sunlei@westlake.edu.cn; wangshaoze@westlake.edu.cn

## Abstract

Two-dimensional conjugated metal–organic frameworks (2D c-MOFs) hold great promise for chemiresistive sensing, electrocatalysis, and energy storage. Their interfacial interactions with metal electrodes can significantly influence electronic properties and device functionality. As a representative 2D c-MOF, $Ni_3(HITP)_2$ (HITP = 2,3,6,7,10,11-hexaiminotriphenylene) exhibits excellent performance in various electronic devices, yet the microscopic mechanism of interfacial interactions in $Ni_3(HITP)_2$/metal heterostructures remains unclear. Here, we combine molecular beam epitaxy synthesis, scanning tunneling microscopic and spectroscopic characterization, and tight-binding analysis to investigate monolayer $Ni_3(HITP)_2$ epitaxially grown on Au(111). The resulting heterostructure forms a commensurate kagome–hexagonal–honeycomb triple-lattice architecture. Energy-dependent local density of states mapping resolves a ligand-derived state at 0.4 eV and dispersive bands dominated by Au surface states near the Fermi level, both of which are reproduced by the tight-binding model. These results demonstrate that the Au(111) substrate induces charge transfer, which in turn leads to Fermi-level pinning, while substrate-mediated hopping reconstructs the band structure. This atomic-scale study of $Ni_3(HITP)_2$/Au(111) epitaxial heterostructures elucidates interlayer coupling mechanisms and advances the understanding of MOF/metal interfaces that are integral to electronic and energy-storage devices.

## Introduction

Two-dimensional conjugated metal–organic frameworks (2D c-MOFs) are crystalline layered materials formed via the coordination-driven self-assembly of metal ions and conjugated organic ligands. These frameworks exhibit highly ordered and diverse lattices (e.g. kagome, hexagonal, honeycomb, and Lieb lattices) with in-plane metal–ligand covalency[1]. Compared with conventional inorganic 2D materials, the lattice geometry and electronic band structures of 2D c-MOFs can be rationally designed at the atomic level, leading to the observation of a broad range of electronic behaviors resembling semiconductors, metals, and even superconductors[2–8]. These electronic functionalities have extended the applications of 2D c-MOFs beyond traditional areas such as gas storage and separation, towards electronic and energy devices spanning field-effect transistors, chemiresistive sensors, supercapacitors, and electrocatalysis[9–16]. In many of these applications, interfaces between frameworks and electrode materials play a key role. The formation of MOF/metal heterostructures can profoundly modify the electron behaviors of the constituent materials through interfacial charge transfer, lattice symmetry reconstruction, and periodic modulation of charge density, which reconfigures Coulombic and magnetic interactions at distinct atomic sites[17–21]. This reconfiguration can significantly renormalize the electronic band structure, potentially altering charge transport and magnetic properties of the MOF and ultimately influencing its performance in electronic devices. Nevertheless, these interfacial phenomena remain largely unexplored. Macroscopic transport measurements typically capture only the ensemble-averaged electron behaviors, offering limited insight into microscopic mechanisms. To this end, the nanoscale construction of MOF/metal heterostructures provides an ideal platform for probing interfacial interactions and microscopic investigations of electronic band structures[22].

$M_3(HITP)_2$ (M = Co, Ni, Cu; HITP = 2,3,6,7,10,11-hexaiminotriphenylene) stands as a promising material platform for investigating 2D c-MOF/metal heterostructures, owing to its well-defined lattice symmetry, excellent electrical properties, and emerging relevance as a quantum material[9,23–25]. The framework is formed through the coordination of square-planar metal ions with tridentate conjugated ligands, yielding a layered structure that coherently integrates three distinct lattices: a metal-centered kagome lattice, a ligand-derived honeycomb lattice, and a hexagonal lattice of vacant pores (Fig. 2). This structural motif is difficult to accomplish in conventional 2D materials. This unique architecture has been predicted to host exotic quantum phenomena including nontrivial band topology and quantum anomalous Hall effect[25–29]. In addition, $M_3(HITP)_2$ exhibits relatively high room-temperature electrical conductivity, which has been attributed to intralayer π–d conjugation and interlayer π–π stacking[24,30,31]. The synergy between electrical conductivity and inherent microporosity has enabled $M_3(HITP)_2$ to achieve high performance in applications ranging from supercapacitors and electrocatalysis to chemiresistive sensing[9,15,32].

Previous studies have demonstrated the ultrahigh-vacuum (UHV) synthesis of $M_3(HITP)_2$ on Au(111) or Cu(111), enabling structural analysis and spectroscopic characterization of the as-grown crystallites[33–35]. These experimental efforts, corroborated by density functional theory (DFT) calculations, have led to the proposal of a metal-centered kagome flat band located at approximately 0.6 eV above the Fermi level[25,35–37]. However, the lack of high-resolution local density of states (LDOS) mapping and comprehensive band structure analysis have hindered unambiguous assignment of orbital characters and band topology. In realistic devices and surface-grown heterostructures, interfacial charge transfer, electronic screening, and substrate-mediated hopping can collectively and substantially reconstruct the electronic band structure of the MOF, potentially shifting its Fermi level and altering electron behaviors[20,22,38–40].

Herein, we investigate monolayer $Ni_3(HITP)_2$ epitaxially grown on Au(111) using scanning tunneling microscopy/spectroscopy (STM/STS) and tight-binding modeling. UHV-based molecular beam epitaxy (MBE) synthesis on Au(111) yielded ordered monolayer crystallites. Energy-resolved LDOS mapping confirmed co-existence of kagome, honeycomb, and hexagonal lattices. Corroborated by tight-binding analysis, they demonstrate the ligand-derived orbitals above the Fermi level as well as pore-confined states and substrate-dominated dispersive bands near the Fermi level. Comparison between the freestanding and renormalized band structures of $Ni_3(HITP)_2$ indicates that electron transfer from Au(111) significantly shifts and pins the Fermi level, while substrate-mediated hopping reconstructs the MOF bands. These results provide a microscopic framework for understanding renormalized electron behaviors at porous MOF/metal interface.

## Synthesis and structural characterization

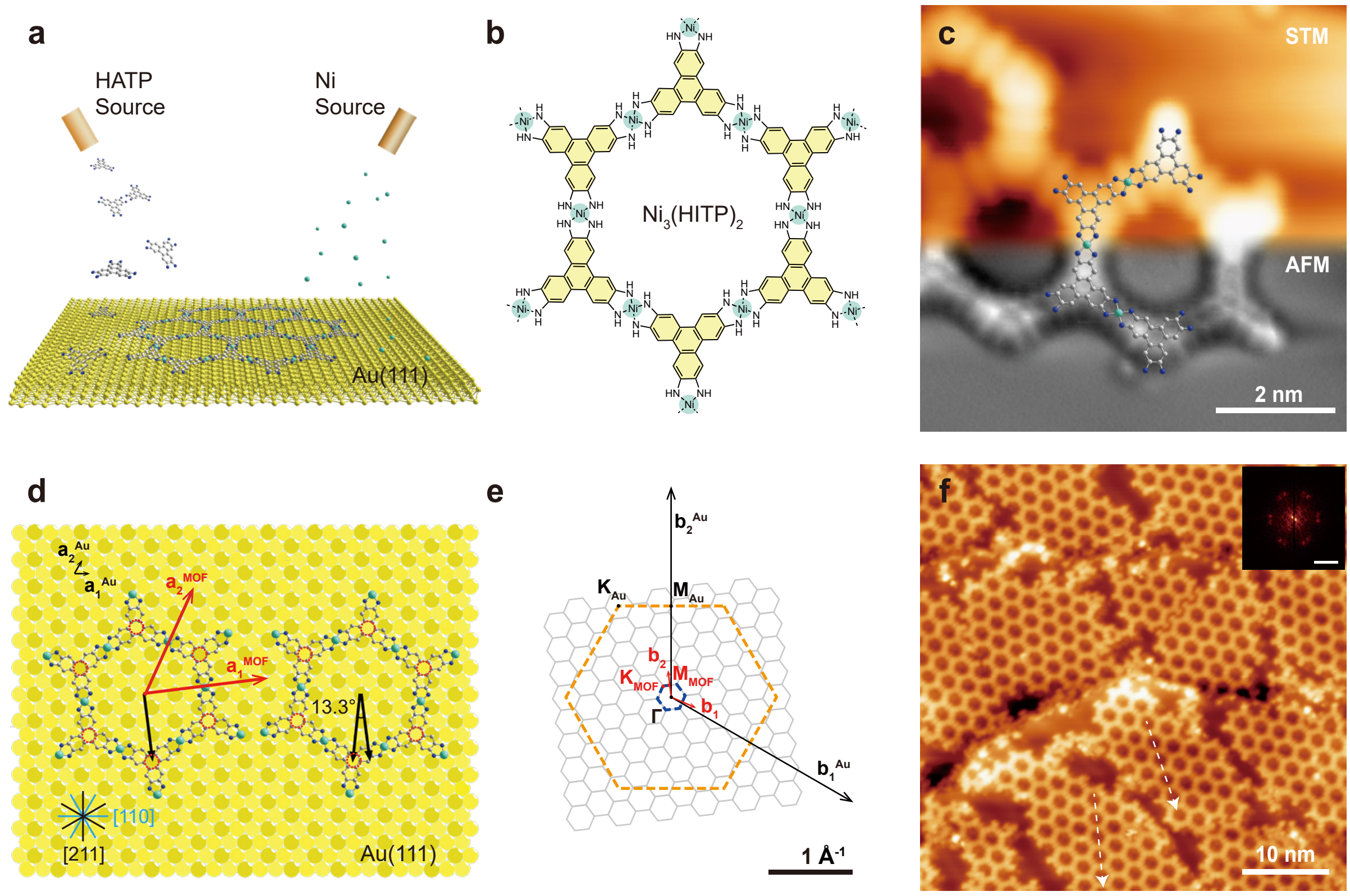

**Fig. 1 | Synthesis and structural characterization of monolayer $Ni_3(HITP)_2$ on Au(111). a**, Schematic of MBE growth of $Ni_3(HITP)_2$. **b**, Structural formula of monolayer $Ni_3(HITP)_2$ featuring a kagome metal lattice (blue) and a honeycomb ligand lattice (yellow). **c**, The STM (top) and q-Plus AFM (bottom) images of $Ni_3(HITP)_2$ acquired in the same region, with corresponding atomic model overlaid. **d**, Two adsorption orientations of $Ni_3(HITP)_2$ on Au(111), rotated by 13.3° relative to each other. Light blue, gray, dark blue, and yellow spheres represent Ni, C, N, and Au, respectively. H atoms are omitted for clarity in **a**, **c**, and **d**. Real-space unit vectors of Au(111) and MOF are marked by black and red arrows. **e**, Brillouin zones of the $Ni_3(HITP)_2$ and the underlying Au(111) substrate in momentum space. They are indicated by blue and orange hexagons with their corresponding reciprocal lattice vectors shown by red and black arrows, respectively. High symmetry points of Au(111) within the first Brillouin zone are marked. **f**, High coverage of $Ni_3(HITP)_2$ domains on Au(111) with two different orientations marked by white dashed arrows confirmed by fast Fourier transform analysis in the inset image. The scale bar in the inset corresponds to 0.5 $nm^{-1}$.

We synthesized $Ni_3(HITP)_2$ by depositing Ni atoms and 2,3,6,7,10,11-hexaaminotriphenylene (HATP) molecules onto atomically clean Au(111) substrate in UHV, followed by annealing at 240 °C, at which Ni deprotonates, reduces, and binds HATP (Fig. 1a, b; see Supporting Information). Q-plus atomic force microscopy (AFM) verified structural integrity of ligands and revealed square-planar $[NiN_4]$ building blocks formed by two HITP (deprotonated HATP) ligands coordinating to each metal ion (Fig. 1c). STM topographic imaging in the same region corroborated this structure. The HITP ligand exhibits an equilaterally triangular electron cloud that is rotated by 60º with respect to its atomic structure[33]. Electron cloud of the metal ion is squeezed into an oblate spherical orbital[33,41]. The self-assembly of these two constituents generates an ordered network. They delineate hexagonal pores that form a hexagonal lattice whose unit cell parameter (namely distance between center of adjacent pores) is 2.2 nm.

In regions on the hexagonal Au(111) surface with a high coverage of $Ni_3(HITP)_2$, crystalline domains adopt two orientations where their symmetry axes rotate by 13.3° with respect to each other (Fig. 1f). The lateral size of each domain is primarily 10 − 20 nm. This phenomenon can be attributed to an adsorption model where the central benzene ring of each HITP ligand is positioned precisely atop a gold atom (Fig. 1d). This configuration corresponds to the energetically favorable atomic site for the extended π-conjugated system under van der Waals interactions with the substrate[33]. This model indicates the formation of $Ni_3(HITP)_2$/Au(111) commensurate heterostructures and yields two lowest-energy adsorption configurations that are oriented at an angle of 13.3° relative to each other, consistent with the experimental observation. Such flexible crystallographic orientations prohibit the growth of large-area single crystals at the atomic level. Owing to the large lattice constant of $Ni_3(HITP)_2$ compared with Au(111), the 1st Brillouin zone of the MOF is substantially reduced and repeated within the reciprocal space of the substrate (Fig. 1e).

**Renormalized triple-lattice system**

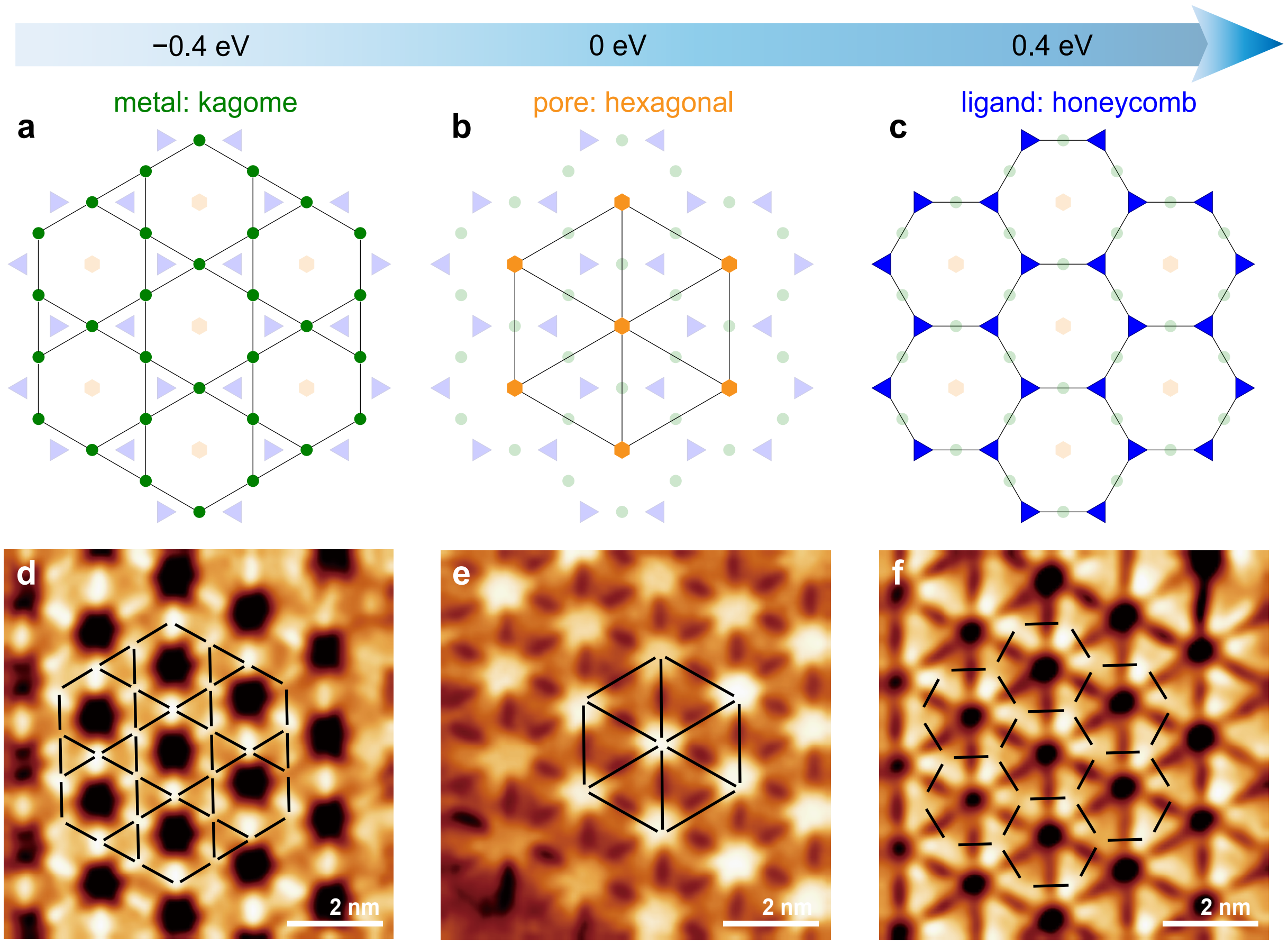

**Fig. 2 | Triple-lattice architecture of monolayer $Ni_3(HITP)_2$ on Au(111). a,** Schematic of kagome lattice formed by metal ions (green), **b,** hexagonal lattice formed by electronic states in pores (orange), **c,** honeycomb lattice formed by HITP ligands (blue). **d,** Experimental STS mapping (I channel, $V_s$ = −0.4 V) highlighting the kagome lattice of metal nodes. **e,** STS mapping (dI/dV channel, $V_s$ = 0 V) revealing the hexagonal pore lattice. **f,** STS mapping (dI/dV channel, $V_s$ = 0.4 V) showing the honeycomb ligand lattice. All STS images were acquired with $I_t$ = 100 pA at 12.5 K. $V_s$: voltage of sample; $I_t$: current of tip.

Unlike its freestanding counterpart whose band structure has been discussed thoroughly in theoretical calculations[25,35–37], the electronic features of $Ni_3(HITP)_2$ on Au(111) are substantially influenced by the substrate. Metal ions form a kagome lattice, HITP ligands constitute a honeycomb lattice, and the periodic pores give rise to a third hexagonal lattice composed of pore-confined electronic states (Fig. 2a–c). The co-existence of these three lattices can be directly visualized by energy-dependent STS mappings (Fig. 2d–f). Consequently, the electronic structure of $Ni_3(HITP)_2$/Au(111) should be treated as a kagome–hexagonal–honeycomb triple-lattice system in which the electron hopping behaviors are fundamentally renormalized. The emergence of pore-confined states overlapping with the rest two lattices introduces additional hopping pathways beyond ideal kagome lattices and provides a versatile platform for band reconstruction at the interface.

## STS mappings and tight binding calculations

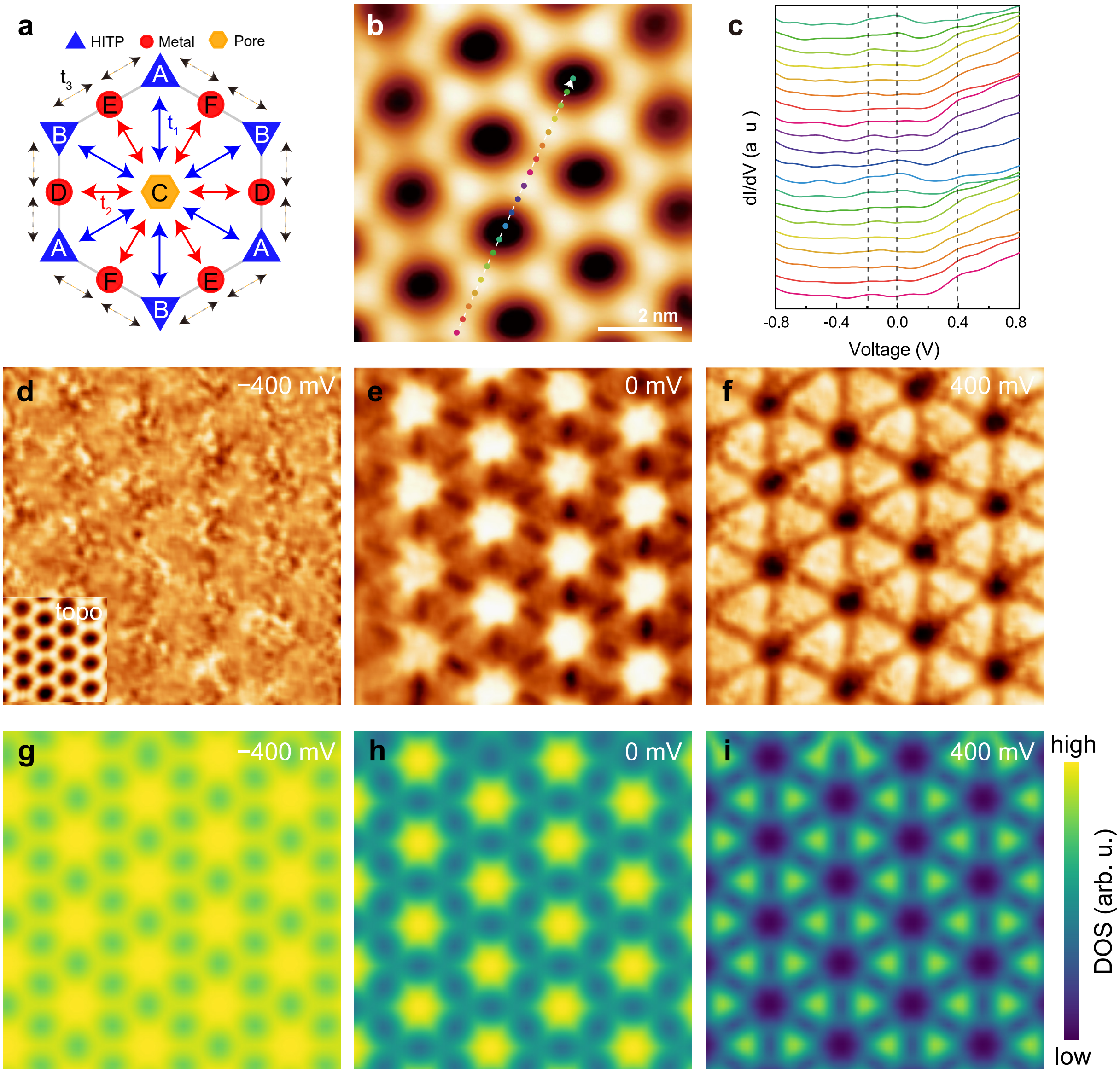

**Fig. 3 | LDOS mapping and tight-binding calculations of $Ni_3(HITP)_2$/Au(111). a**, Proposed hopping model for the kagome–hexagonal–honeycomb triple-lattice system. Blue triangles, red circles, and yellow hexagons represent HITP ligands, metal ions, and confined states in the pores, respectively. $t_1$, $t_2$, and $t_3$ represent hopping constants between different sites. **b**, The

topography of measured area with color-coded dots along the dashed white arrow. **c**, Spatially resolved dI/dV spectra acquired along the color-coded path marked in **b**. DOS peaks are marked by dash lines. **d−f**, Experimental STS mappings of $Ni_3(HITP)_2$ at −400 mV, 0 mV, and 400 mV, with corresponding topographic image shown as an inset in **d**. STS measurements were taken with $V_s = -1$ V, $I_t = 100$ pA. **g−i**, Tight binding calculations of LDOS mappings at −400 mV, 0 mV, and 400 mV.

We experimentally examined the LDOS by taking a series of dI/dV spectra across a representative $Ni_3(HITP)_2$ domain along the line crossing Ni atoms, HITP ligands, and pore centers (Fig. 3b, c). Three characteristic DOS peaks were observed at approximately −0.2 eV, 0 eV, and 0.4 eV. The peak at 0.4 eV appears mainly atop MOFs while becoming absent inside pores. In contrast, the features at −0.2 eV and 0 eV are most prominent inside the pores and become weaker atop the MOF. These observations indicate different origins of electronic states from both pore-confinement effect and orbitals of MOF at different atomic sites. To visualize DOS distributions at various energies, we conducted STS mappings of one square region at −0.4 eV, 0 eV and 0.4 eV (Fig. 3d−f). The 2D DOS at 0.4 eV exhibits a honeycomb network of triangles coinciding with HITP ligands, demonstrating the ligand origin of electronic bands above the Fermi level that is inconsistent with previous assignments[37, 41]. At 0 eV, stronger pore-confined states emerge and couple with weakened HITP molecular orbitals. It implies that surface electrons of Au(111) are scattered and trapped inside pores of $Ni_3(HITP)_2$, reconstructing the electron hopping modes and dominating the electrical conduction behaviors at the interface. However, the STS mapping at −0.4 eV revealed comparable contributions from Ni atoms, HITP ligands and pore-confined states resulting in featureless DOS distributions. These STS mappings indicate negligible DOS from Ni as well as weak metal–ligand $\pi$–d conjugation near the Fermi level at the $Ni_3(HITP)_2$/Au(111) interface.

In order to understand its electron behavior, we conducted tight-binding analysis to explore the electronic band structure near the Fermi level (−1 eV – 1 eV) of the $Ni_3(HITP)_2$/Au(111) epitaxial heterostructure. We propose that the MOF-modulated surface states of Au(111) play a key role in the band structure—they mediate electron hopping between different sites in $Ni_3(HITP)_2$, yielding stronger hopping than that within the framework, and couple with two pore-confined states at −0.2 eV and 0 eV. The overall kagome, hexagonal, honeycomb triple lattices of the heterostructure constitute an interlocked 8-band system with the following Hamiltonian (see details in the Supporting Information):

$$H_{NN} = -V(C_1^\dagger C_{Au} + C_2^\dagger C_{Au}) - t_1(A^\dagger C_{Au} + B^\dagger C_{Au}) - t_2(D^\dagger C_{Au} + E^\dagger C_{Au} + F^\dagger C_{Au}) - t_3(D^\dagger A + E^\dagger A + F^\dagger A) - t_3(D^\dagger B + E^\dagger B + F^\dagger B) + h.c. \quad (1)$$

This electron system comprises of a gold surface state with parabolic dispersion ($C_{Au}^\dagger$), two pore-confined states ($C_1^\dagger$ and $C_2^\dagger$), two HITP sites ($A^\dagger$ and $B^\dagger$), and three Ni sites ($D^\dagger$, $E^\dagger$, and $F^\dagger$). The surface state of Au(111) was assigned to be a k-dependent parabola above −0.45 eV[42], the energy of ligand sites were assigned as 0.4 eV based on the STS mapping results, whereas that of Ni sites were assigned arbitrarily to be −2.2 eV such that they do not contribute to the bands near the Fermi level. Hopping constants $t_1$ and $t_2$ were assumed to be much larger than $t_3$ and $V$, where $V$ describes the onsite coupling between the surface state of Au(111) and pore-confined states. Only the nearest neighbor hopping was considered (Fig. 3a).

The above model yields STS mapping simulations (Fig. 3g–i) that match experimental observations of the $Ni_3(HITP)_2$/Au(111) heterostructure (Fig. 3d–f). Specifically, simulated images reproduce triangular HITP states at 0.4 eV and hexagram Au(111) states at the Fermi level. As gold, HITP, and metal ions contribute comparable LDOS at −0.4 eV, the simulated image at this energy shows a low contrast, which explains the featureless experimental image. Such consistency between simulated and experimental STS mapping confirms the validity of our model. It implies that the electrical conduction in the $Ni_3(HITP)_2$/Au(111) heterostructure is predominantly governed by electrons from Au(111) and HITP—insufficient orbital overlap between metal ions and ligands near the Fermi level prevents efficient electron hopping within the framework.

**Band renormalization analysis**

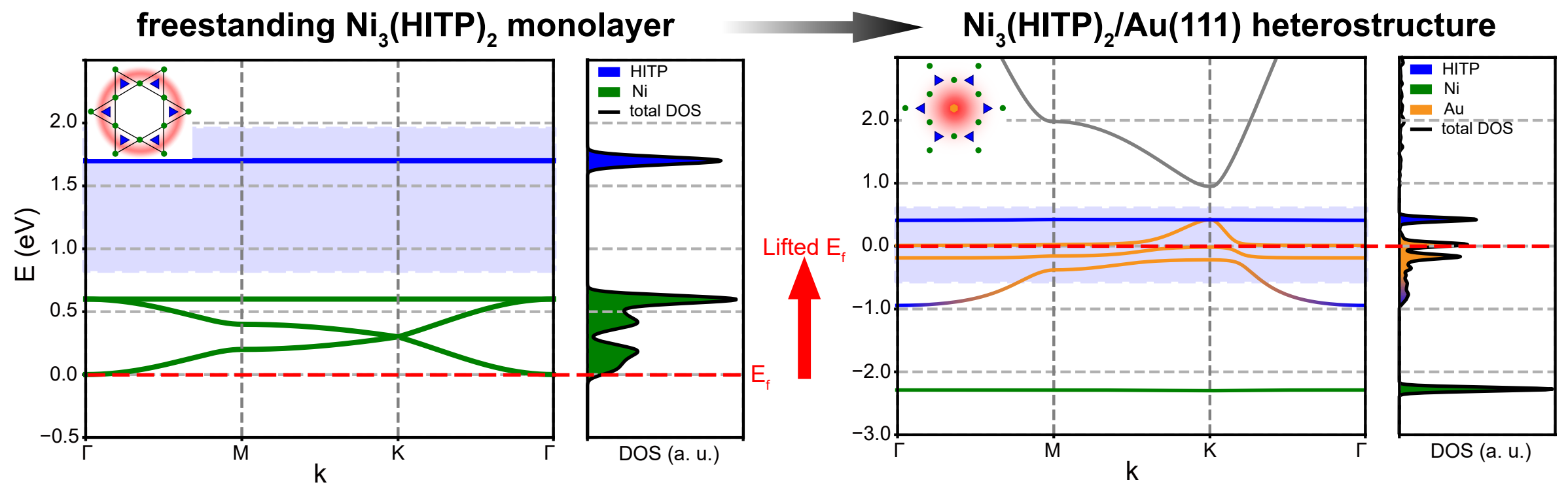

**Fig. 4 | Substrate-induced band renormalization in $Ni_3(HITP)_2$/Au(111).** Band structure transition from the freestanding $Ni_3(HITP)_2$ monolayer (left) to the $Ni_3(HITP)_2$/Au(111) heterostructure (right). The band structure on the left was calculated by tight binding model to simulate reported DFT calculation results comprising of a kagome flat band at 0.6 eV together with separated HITP orbitals at about 1.7 eV[35,36]. The band structure on the right was plotted according to the newly proposed triple-lattice hopping model (Fig. 3a), which incorporates Au(111) surface states, pore-confined states, Fermi-level pinning, and substrate-mediated hopping.

To understand the reconstructed electronic structure, we compared the band structures of a freestanding $Ni_3(HITP)_2$ monolayer and the $Ni_3(HITP)_2$/Au(111) heterostructure (Fig. 4). In the former, a kagome-derived flat band appears at 0.6 eV above the Fermi level together with two dispersive crossing bands, while the ligand-derived isolated molecular orbitals form another flat band at higher energies (1.7 eV). They reproduce the characteristic band features calculated by previous DFT analysis[35,36]. In the heterostructure, the work-function difference between $Ni_3(HITP)_2$ and Au(111) drives substantial electron transfer from the substrate to the MOF, which pins the Fermi level and shifts it upward. The additional substrate-mediated hopping further strongly reconstructs the original dispersions. Such substrate-induced Fermi level pinning and band renormalization account for the absence of appreciable Ni-derived LDOS from −0.2 eV to 0.6 eV as well as two overlapped Ni-derived flat bands at −2.2 eV from renormalized Ni-based kagome lattice. Phase cancellations of asymmetric wave functions from two different HITP sites in the honeycomb lattice inhibit electron hopping towards the hexagonal lattice (Fig. S3), therefore inducing a ligand-centered lattice-frustrated flat band at 0.4 eV that is different from previous assignment[37, 41]. Perturbations from weak coupling with adjacent Ni sites induce a tiny dispersion of this band.

Near the Fermi level, our model yields three dispersive bands between −1.0 eV and 0.4 eV within the first Brillouin zone. Their LDOS are dominated by electronic states of Au(111) with slight contributions from both Ni and HITP (Fig. S2b). The coupling between the surface state and pore-confined states generates small energy gaps between adjacent bands. The band dispersion is inherited from the surface state, which endorse the $Ni_3(HITP)_2$/Au(111) heterostructure with metallicity. Consequently, the electronic states near the Fermi level are dominated by Au(111), whereas the MOF supplies the periodic microporous potential that confines and modulates the substrate surface states. Band renormalization therefore occurs in both components of the $Ni_3(HITP)_2$/Au(111) heterostructure.

**Discussion**

The foregoing results demonstrate reciprocal band renormalization in both monolayer $Ni_3(HITP)_2$ and the Au(111) surface upon formation of the epitaxial heterostructure. LDOS mappings and tight binding analysis reveal direct evidence of interfacial charge transfer and the key role of Au(111) surface states in charge hopping, which pins the Fermi level of $Ni_3(HITP)_2$ and renormalizes its band structure. As a result, the Au(111) surface states dominate at the Fermi level and ligand-derived states with lattice frustration give rise to a flat band at 0.4 eV. The latter observation corrects the previous attribution to the Ni-based kagome lattice[35] and highlights the non-negligible influence of substrate to the adsorbed MOF. This result is inconsistent with DFT-predicted band structures of the freestanding $Ni_3(HITP)_2$ monolayer and its heterostructure with Au(111)[25,35,37]. Although DFT calculations are widely used to understand electron behaviors in MOF/metal interfaces[43–45], our work shows its limitation for analyzing such complex systems and raises stringent demand of physical insights into the hopping model corroborated by high-quality energy-dependent LDOS mappings.

As the metallic substrate induces significant band renormalization, in order to examine the MOF's intrinsic band structure, it is indispensable to deposit an insulating layer (e.g. hexagonal boron nitride, MgO, NaCl) on the metallic substrate or use semiconducting substrates (e.g. $MoS_2$ and doped Si(111))[22,46,47]. Meanwhile, commensurate lattice matching effectively modulates surface states of Au(111), generating pore-confined states at the interface. These states enhance the LDOS at the Fermi level and promote the accumulation of substrate-derived charges within the pores, thereby inducing periodic potential modulations at the surface that may lead to electron localization and band flattening in the substrate. These effects could enable the rational design of correlated systems and facilitate the optimization of MOF/electrode interfaces in electronic devices by tuning the structural and electronic properties of both constituents, potentially enhancing electrocatalytic activity and chemiresistive sensitivity among others.

**Data availability**

The data that support the findings of this study are available from the corresponding authors upon request.

**Acknowledgments**

This work was financially supported by the Zhejiang Provincial Natural Science Foundation of China (XHD23B0301). MBE and STM experiments were done at Songshan Lake Materials Laboratory. Part of this work made use of the experimental facilities at Shanghai Synchrotron Radiation Facility and was supported by the Strategic Priority Research Program of the Chinese Academy of Science (XDB1180000). We thank Yunhui Wang, Dr. Jiaqi Lin and Dr. Chaozhi Huang for assistance with preliminary experiments at Instrumentation and Service Center for Physical Sciences at Westlake University, Dr. Huiru Liu and Ziheng Zhang for assistance with STM experiments, Prof. Wei Zhu for theoretical suggestions, Prof. Ruihua He's laboratory and Dr. Shaoshan Wang for experimental equipment assistance, Prof. Changxi Zheng for helpful discussions, and Prof. Yi Du for experimental instructions. L.S. thanks Prof. Zhongyue Zhang, Prof. Jin-Hu Dou, Prof. Xiaohui Qiu, Prof. Mengxi Liu, Prof. Carl K. Brozek, Prof. Jie Wu, and Prof. Shuigang Xu for helpful discussions.

**Author contributions**

S.W. and L.S. conceived the idea and designed the experiments. X.Y., S.W., and X.H. conducted the MBE growth of $Ni_3(HITP)_2$ and subsequent STM & q-Plus AFM characterization. Z.S. synthesized HATP molecules. Z.W. and F.S. assisted with deposition of HATP molecules. S.W. performed theoretical calculations. X.Y., S.W. and L.S. analyzed the experimental data and wrote the manuscript. All the authors contributed to the data analysis and provided feedback on the manuscript.

**Competing interests**

The authors declare no competing interests.

**Supporting information**

Materials, synthesis, methods, STM images, and tight-binding analysis. The Supporting information is available free of charge at https://pubs.acs.org.

# Supplementary Information for "Band Renormalization in a Metal–Organic Framework/Au(111) Triple-Lattice Heterostructure"

Xiaoqing Yuan[1,2,3,†], Shaoze Wang[2,3,4,†,*], Xiaoyue He[5], Zhecheng Sun[3,4], Zihao Wang[6], Fei Song[7], Lei Sun[2,3,4,*]

[1]Department of Physics, Fudan University, Shanghai 200433, China

[2]Department of Physics, School of Science and Research Center for Industries of the Future, Westlake University, Hangzhou 310030, Zhejiang Province, China

[3]Institute of Natural Sciences, Westlake Institute for Advanced Study, Hangzhou 310024, Zhejiang Province, China

[4]Department of Chemistry, School of Science and Research Center for Industries of the Future, Westlake University, Hangzhou 310030, Zhejiang Province, China

[5]Songshan Lake Materials Laboratory, Dongguan 523808, Guangdong Province, China

[6]School of Physics, Hunan Key Laboratory for Super-Microstructure and Ultrafast Process, Central South University, Changsha 410083, Hunan Province, China

[7]Shanghai Synchrotron Radiation Facility, Shanghai Advanced Research Institute, Chinese Academy of Sciences, 20100, Shanghai, China

[†]These authors contributed equally to this work.

*Email: sunlei@westlake.edu.cn; wangshaoze@westlake.edu.cn

**$Ni_3(HITP)_2$ synthesis on surface**

All commercially available chemicals were used without further purification unless otherwise noted. Triphenylene, benzophenone imine, 1.1'-Binaphthyl-2.2'-diphemyl phosphine, and tris(dibenzylidenacetone)dipalladium(0), were purchased from Bide Pharmatech Co. Ltd. Iron shavings, sodium tert-butoxide and nitrobenzene were purchased from Macklin. Tetrahydrofuran (THF) and dichlorobenzene were purchased from Innochem. Toluene, diethyl ether, liquid bromine and hydrochloric acid was purchased from Sinopharm Chemical Reagent Co. Ltd. 2,3,6,7,10,11-hexaaminotriphenylene (HATP) was synthesized based on literature procedures[1]. The Au(111) substrate was supplied by Labmates Technology Co. Ltd. Ni source by PrMat Co. Ltd.

Before deposition, HATP powder was degassed at 160 °C for 1–2 hours in a separate chamber until the pressure stabilized at approximately $10^{-6}$ mbar, and the Ni source was heated to approximately 1250 °C. An atomically clean Au(111) surface was prepared by repeated cycles of argon ion sputtering and annealing at 540 °C. Surface cleanliness was confirmed by STM through observation of the characteristic herringbone reconstruction and monoatomic steps of Au(111) (Fig. S1a). For $Ni_3(HITP)_2$ growth, Ni was first deposited onto the Au(111) substrate held at room temperature for 30 s under a background pressure of $10^{-9}$ mbar, resulting in a sub-monolayer Ni coverage exceeding 30% (Fig. S1b). HATP molecules were then evaporated from a crucible at 160 °C for 20 s onto the substrate held at room temperature in a separate chamber at $10^{-6}$ mbar (Fig. S1c). The sample was subsequently annealed at 240 °C for 15 minutes under UHV ($10^{-10}$ mbar) to form large-area, polycrystalline, monolayer $Ni_3(HITP)_2$ films (Fig. S1d).

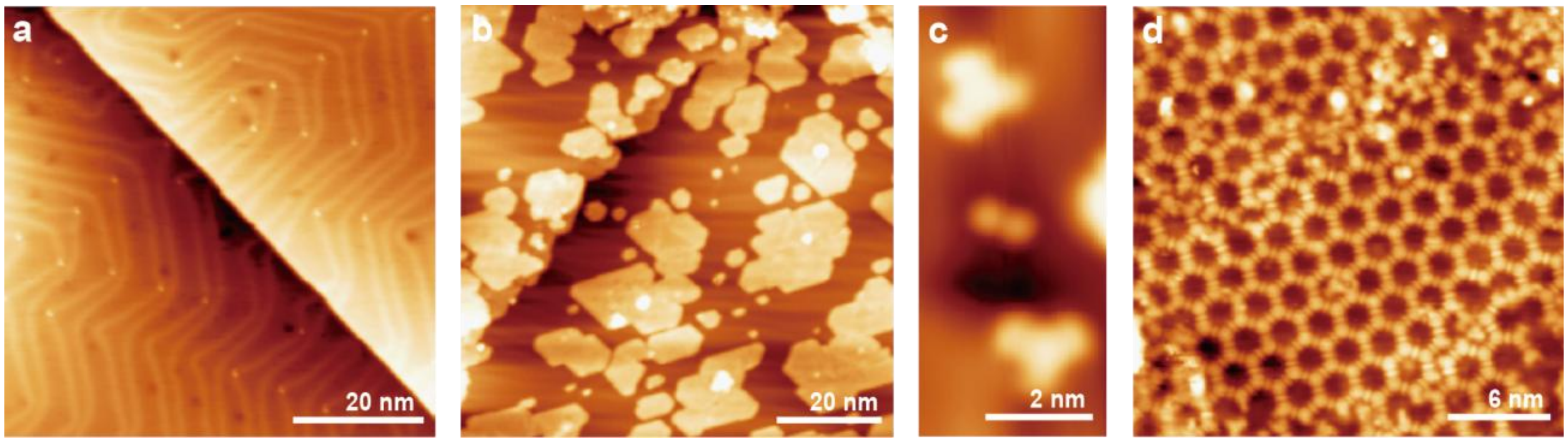

**Fig. S1. STM images of $Ni_3(HITP)_2$ growth process. a,** Atomically clean Au(111) surface before deposition. **b,** Sub-monolayer coverage of Ni islands deposited onto Au(111). **c,** Isolated HATP molecules on Au(111). **d,** The domains of $Ni_3(HITP)_2$ after the growth process.

## STM/ q-Plus AFM measurements

All STM and spectroscopic measurements were performed using a commercial low-temperature STM (Infinity STM, Omicron) at 12.5 K under a base pressure of approximately 7.0 × $10^{-11}$ mbar. Topographic images, STS line spectra, and mappings were acquired in constant-current mode using a setpoint of $V_s = -1$ V and $I_t = 100$ pA. The dI/dV spectra were measured using lock-in detection with an AC modulation voltage of 25 mV at a frequency of 790 Hz superimposed on the bias voltage. The STM tip condition was verified by observation of the surface state of clean Au(111).

Q-Plus AFM measurements were performed using an Omicron Infinity low-temperature STM at 12.5 K under a base pressure of $10^{-11}$ mbar. A CO-functionalized tip was prepared by picking up an adsorbed CO molecule from the Au(111) substrate while scanning with a setpoint of $V_s$ = 100 mV and $I_t$ = 100 pA. Successful tip functionalization was confirmed by imaging individual CO molecules with altered topographic contrast.

All STM and AFM images were processed using WSxM software[2]. Spectra were appropriately smoothed to remove background thermal noise.

## Triple lattice hopping model

A simplified hopping model of the $Ni_3(HITP)_2$ triple lattices is used to explain the observed experimental STS measurements. HITP ligands, metal ions and trapped states inside MOF pores constitute the honeycomb, kagome and hexagon lattices respectively. HITP ligands have two different environmental sites named with sub-lattice A and B. Metal ions have three sub-lattices D, E and F. While two resonant states and one Au(111) surface state inside pores are located at lattice C. The Hamiltonian of the system is defined as:

$$H = H_0 + H_{NN} \tag{10}$$

$$H_0 = e_0 A^\dagger A + e_0 B^\dagger B + e_1 {C_1}^\dagger C_1 + e_2 {C_2}^\dagger C_2 + e_3 D^\dagger D + e_3 E^\dagger E + e_3 F^\dagger F + e_{Au} {C_{Au}}^\dagger C_{Au} \tag{11}$$

$$H_{NN} = -t_1 \sum_{\vec{R},\vec{\delta}} [A^{\dagger}(\vec{R}+\vec{\delta}_1)C_{Au}(\vec{R}) + B^{\dagger}(\vec{R}+\vec{\delta}_2)C_{Au}(\vec{R})]$$
$$- V \sum_{\vec{R},\vec{\delta}} [{C_1}^{\dagger}(\vec{R})C_{Au}(\vec{R}) + {C_2}^{\dagger}(\vec{R})C_{Au}(\vec{R})]$$
$$- t_2 \sum_{\vec{R},\vec{\delta}} [D^{\dagger}(\vec{R}+\vec{\delta}_3)C_{Au}(\vec{R}) + E^{\dagger}(\vec{R}+\vec{\delta}_4)C_{Au}(\vec{R}) + F^{\dagger}(\vec{R}+\vec{\delta}_5)C_{Au}(\vec{R})]$$
$$- t_3 \sum_{\vec{R},\vec{\delta}} [D^{\dagger}(\vec{R}+\vec{\delta}_6)A(\vec{R}) + E^{\dagger}(\vec{R}+\vec{\delta}_7)A(\vec{R}) + F^{\dagger}(\vec{R}+\vec{\delta}_8)A(\vec{R})]$$
$$- t_3 \sum_{\vec{R},\vec{\delta}} [D^{\dagger}(\vec{R}+\vec{\delta}_9)B(\vec{R}) + E^{\dagger}(\vec{R}+\vec{\delta}_{10})B(\vec{R}) + F^{\dagger}(\vec{R}+\vec{\delta}_{11})B(\vec{R})] + h.c. \tag{12}$$

$H_0$ represents the different onsite energies of corresponding sub-lattices $e_0 = 0.4$ eV, $e_1 = -0.2$ eV, $e_2 = 0$ eV, $e_3 = -2.2$ eV together with k-dependent energy $e_{Au}$ of Au(111) parabolic band. −0.2 eV and 0 eV are set as energies of two resonant states only coupled with Au(111) surface state. 0.4 eV is defined due to the observed localized molecular orbitals. $H_{NN}$ defines the nearest neighbor hopping Hamiltonian among A, B, C lattice sites as well as couplings between two resonant states $C_1$, $C_2$ and Au(111) surface state electrons inside pores. Operator $A^{\dagger}(\vec{R})$ $(A(\vec{R}))$ means the creation (annihilation) of electrons at position $\vec{R}$ in sub-lattice A and h.c. refers to their corresponding Hermitian conjugate terms. $\vec{\delta}_i$ is the vector directing to the nearest neighboring sub-lattice site. In such setup, only electrons from sub-lattice C can hop to elsewhere and vice versa. While sub-lattice A and B cannot directly communicate with each other because they have to pass through sub-lattice D, E and F. The same works for the hopping of sub-lattice D, E and F towards elsewhere.

Fourier expanding the Hamiltonian $H_{NN}$ into k space results with:

$$H_{NN} = -t_1 \sum_{\vec{\delta}_1,\vec{\delta}_2} \int_{-\infty}^{+\infty} [e^{-i\vec{k}\cdot\vec{\delta}_1} A^{\dagger}(\vec{k})C_{Au}(\vec{k}) + e^{-i\vec{k}\cdot\vec{\delta}_2} B^{\dagger}(\vec{k})C_{Au}(\vec{k})]d\vec{k}$$
$$- V \int_{-\infty}^{+\infty} [{C_1}^{\dagger}(\vec{k})C_{Au}(\vec{k}) + {C_2}^{\dagger}(\vec{k})C_{Au}(\vec{k})]d\vec{k}$$
$$-t_2 \sum_{\vec{\delta}_3,\vec{\delta}_4,\vec{\delta}_5} \int_{-\infty}^{+\infty} [e^{-i\vec{k}\cdot\vec{\delta}_3} D^{\dagger}(\vec{k})C_{Au}(\vec{k}) + e^{-i\vec{k}\cdot\vec{\delta}_4} E^{\dagger}(\vec{k})C_{Au}(\vec{k}) + e^{-i\vec{k}\cdot\vec{\delta}_5} F^{\dagger}(\vec{k})C_{Au}(\vec{k})]d\vec{k}$$

$$-t_3 \sum_{\vec{\delta}_6,\vec{\delta}_7,\vec{\delta}_8} \int_{-\infty}^{+\infty} [e^{-i\vec{k}\cdot\vec{\delta}_6} D^\dagger(\vec{k})A(\vec{k}) + e^{-i\vec{k}\cdot\vec{\delta}_7} E^\dagger(\vec{k})A(\vec{k}) + e^{-i\vec{k}\cdot\vec{\delta}_8} F^\dagger(\vec{k})A(\vec{k})]d\vec{k}$$

$$-t_3 \sum_{\vec{\delta}_9,\vec{\delta}_{10},\vec{\delta}_{11}} \int_{-\infty}^{+\infty} [e^{-i\vec{k}\cdot\vec{\delta}_9} D^\dagger(\vec{k})B(\vec{k}) + e^{-i\vec{k}\cdot\vec{\delta}_{10}} E^\dagger(\vec{k})B(\vec{k}) +$$

$$e^{-i\vec{k}\cdot\vec{\delta}_{11}} F^\dagger(\vec{k})B(\vec{k})]d\vec{k} + h.c.$$

(13)

The vectors $\vec{\delta}_{1-11}$ define the displacements from the origin sub-lattice C (0, 0) to nearest neighbor sites A, B, D, E, F; from A to D, E, F and from B to D, E, F respectively. The distance from C to A is used as the lattice constant $a = 13$ Å which derives from the STM experimental measurement. The hopping parameter $t_1$ and $t_2$ are set as 1 and $t_3$ as 0.1 in order to reproduce the strong coupling strength between Au(111) surface state electrons and the rest as well as the weak couplings between HITP ligands and metal ions near the Fermi level. V = 0.2 defines the weak couplings between resonant states and Au(111) surface state electrons inside pores.

Projecting the Hamiltonian (Equation 10) onto each sublattice orbital in Hilbert space leads to the matrix form:

$$H = \begin{bmatrix} e_0 & 0 & 0 & 0 & f_6^* & f_7^* & f_8^* & f_1 \\ 0 & e_0 & 0 & 0 & f_9^* & f_{10}^* & f_{11}^* & f_2 \\ 0 & 0 & e_1 & 0 & 0 & 0 & 0 & V \\ 0 & 0 & 0 & e_2 & 0 & 0 & 0 & V \\ f_6 & f_9 & 0 & 0 & e_3 & 0 & 0 & f_3 \\ f_7 & f_{10} & 0 & 0 & 0 & e_3 & 0 & f_4 \\ f_8 & f_{11} & 0 & 0 & 0 & 0 & e_3 & f_5 \\ f_1^* & f_2^* & V & V & f_3^* & f_4^* & f_5^* & e_{Au} \end{bmatrix} \quad (14)$$

with

$$f_i = \begin{cases} -t_{1,2} \sum_{\vec{\delta}_i} e^{-i\vec{k}\cdot\vec{\delta}_i}, \ i \le 5 \\ -t_3 \sum_{\vec{\delta}_i} e^{-i\vec{k}\cdot\vec{\delta}_i}, \ i \ge 6 \end{cases} \quad (15)$$

Solving the eigenvalue of the Hamiltonian matrix (Equation 14) renders the E-k dispersion relation plotted along high symmetry points (Fig. S2a). The corresponding average orbital contributions of HITP, metal and pore-trapped states demonstrate their relative weights at varied energies from −6 eV to 4 eV (Fig. S2b). The LDOS mappings (Fig. 3g–i) are simulated based on the square of wave functions. The orbital weights are assigned according to their distances to the interested atomic site in a Gaussian distribution. The closest orbital contributes to the most. Specifically, the molecular orbitals of HITP are formulated into a triangular shape in order to match our experimental observations. In order to match image contrast from experimental observations, the weight factor of Au(111) surface states is enhanced while that of resonant states is diminished.

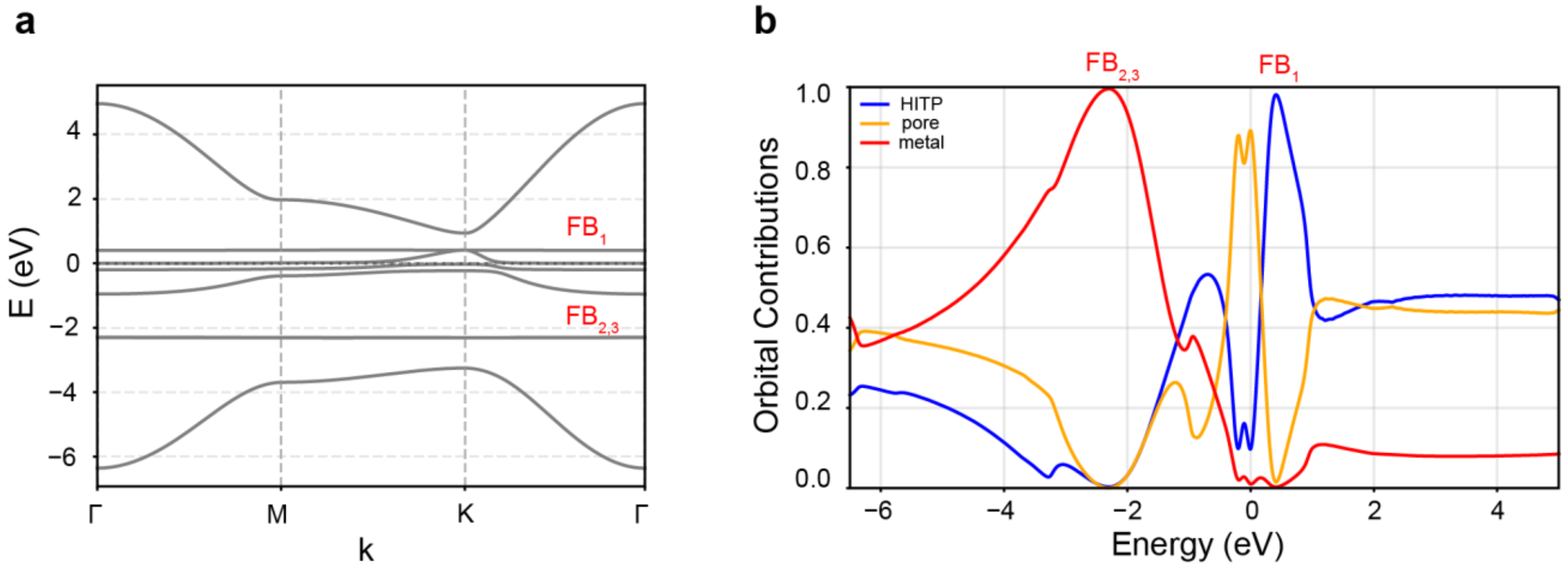

**Fig. S2. Tight binding analysis of monolayer $Ni_3(HITP)_2$/Au(111) triple lattices. a,** Band structure of $Ni_3(HITP)_2$/Au(111) along high symmetry points. FB indicates flat band. **b,** Normalized orbital contributions from HITP, pore and kagome metal sites at different energies.

**Flat band analysis at 0.4 eV**

The weights of wave functions of sub-lattice A (Fig. S3a) and B (Fig. S3b) in this band are almost 0.5 respectively. It means that orbital contributions of the flat band at 0.4 eV come from the HITP ligands. Phase difference with values of either $+\pi$ or $-\pi$ between sub-lattice A and B distributes in an asymmetric way within the 1$^{st}$ Brillouin zone boundary (Fig. S3c). It leads to the phase cancellations of wave functions of sub-lattice A and B which inherently inhibit the electron hopping towards sub-lattice C. Weak couplings (considering the small hopping parameter $t_2$ large onsite energy difference of 0.4 eV and −2.2 eV) with kagome sub-lattice D, E, and F make the flat band slightly dispersive. As a result, electron localizations at triangular molecular orbitals of HITP prohibit the kinetic energies from diffusing to the rest. With specific bottom-up design of magnetic properties of coordinating ligands, electron localization may induce strong correlations at the spin center. Spin orbit couplings will open the energy gap between the flat band and the dispersive band near the Fermi level (whose band touching happens at K points). Topological edge states at the molecular orbitals may appear in an appropriate condition.

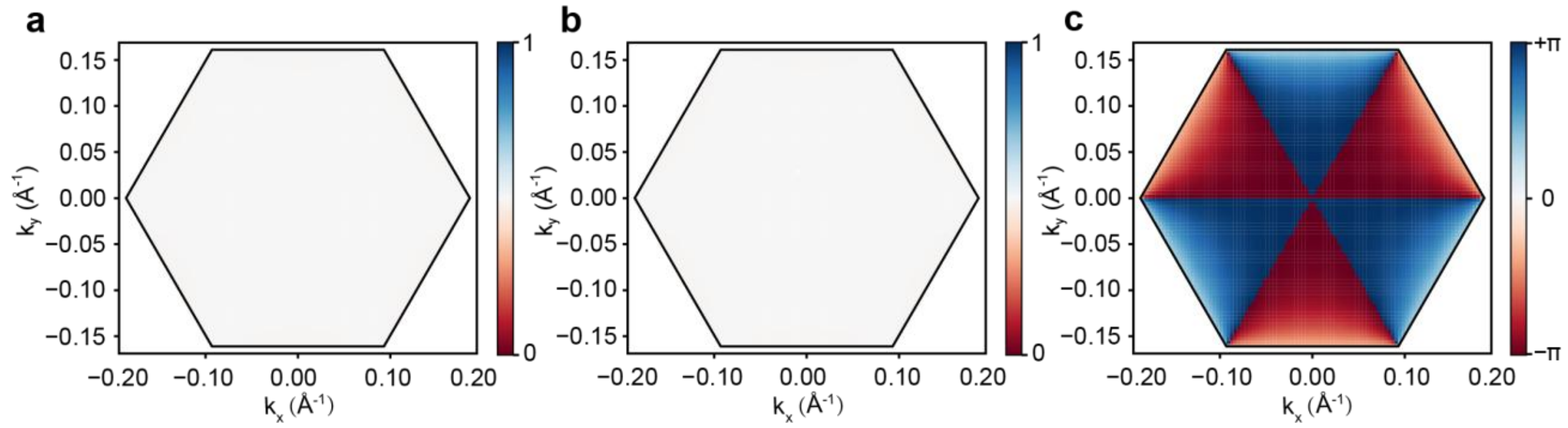

**Fig. S3. Wave functions analysis of honeycomb sub-lattice A and B. a,** Weight distributions of sub-lattice A wave function within the 1st Brillouin zone. **b,** Weight distributions of sub-lattice B wave function within the 1st Brillouin zone. **c,** Phase difference of wave functions within sub-lattice A and B.